\documentclass[final,5p,times,twocolumn]{elsarticle}

\usepackage{graphicx}

\journal{Nuclear Instruments Method}
\begin{document}
\begin{frontmatter}
\title{HD gas analysis with Gas Chromatography and Quadrupole Mass Spectrometer}
\author[a]{T.~Ohta\corref{cor1}}
\cortext[cor1]{Corresponding author.}
\ead{takeshi@rcnp.osaka-u.ac.jp}

\author{S.~Bouchigny$^{b,c}$, J.-P.~Didelez$^{b}$, M.~Fujiwara$^{a}$, K.~Fukuda$^{d}$, H.~Kohri$^{a}$, T.~Kunimatsu$^{a}$, C.~Morisaki$^{a}$, S.~Ono$^{a}$, G.~Rouill\'e$^{b}$, M.~Tanaka$^{e}$, K.~Ueda$^{a}$, M.~Uraki$^{a}$, M.~Utsuro$^{a}$, S.Y.~Wang$^{f,g}$, and M.~Yosoi$^{a}$}
\address[a]{ Research Center for Nuclear Physics, Osaka University, Mihogaoka 10-1, Ibaraki, Osaka 567-0047, Japan}
\address[b]{IN2P3, Institut de Physique Nucl\'{e}aire, F-91406 ORSAY, France}
\address[c]{CEA LIST, BP6-92265  Fontenay-aux-Roses, CEDEX, France}
\address[d]{Kansai University of Nursing and Health Sciences, Shizuki Awaji 656-2131, Japan}
\address[e]{Kobe Tokiwa University, Ohtani-cho 2-6-2, Nagata, Kobe 653-0838, Japan}
\address[f]{Institute of Physics, Academia Sinica, Taipei 11529, Taiwan}
\address[g]{Department of Physics, National Kaohsiung Normal University, Kaohsiung 824, Taiwan}

\begin{abstract}
A gas analyzer system has been developed to analyze Hydrogen-Deuteride (HD) gas for producing frozen-spin polarized HD targets, which are used for hadron photoproduction experiments at
SPring-8.
Small amounts of ortho-H$_{2}$ and para-D$_{2}$ gas mixtures
($\sim$0.01\%) in the purified HD gas
are a key to realize a frozen-spin polarized target.
However, there was an intrinsic difficulty to measure these small
mixtures in the HD gas with a quadrupole mass spectrometer (QMS) because D$^{+}$ and
[H$_{2}$D]$^{+}$ produced from the ionization of HD molecules were misidentified as H$_{2}$ and
D$_{2}$ molecules, respectively, and became backgrounds for the measurement of
the H$_{2}$ and D$_{2}$ concentrations.
In addition, the ortho-H$_{2}$ and para-D$_{2}$ are not distinguished from the para-H$_{2}$ and ortho-D$_{2}$, respectively, with the QMS. 
In order to obtain reliable concentrations of these gas mixtures in the
HD gas, we produced a new gas analyzer system combining two independent measurements with the gas chromatography and the QMS.
Helium or neon gas was used as a carrier gas for the gas chromatography which was cooled at $\sim$110 K.
The para-H$_{2}$, ortho-H$_{2}$, HD, and D$_{2}$ are separated using the retention time of the gas chromatography and the mass/charge ratio.
Although the para-D$_{2}$ is not separated from the ortho-D$_{2}$, the total amount of the D$_{2}$ is measured without the [H$_{2}$D]$^{+}$ background.
The ortho-H$_{2}$ concentration is also measured separately from the
D$^{+}$ background.
It is found that the new gas analyzer system can measure small
concentrations of $\sim$0.01\% for the otho-H$_2$ and D$_2$ with good S/N ratios.
\end{abstract}
\end{frontmatter}

\section{Introduction}
\label{intro}
Double polarization measurements for $\phi$ and K meson
photo productions with a polarized target and a polarized
photon beam are a sensitive means  to investigate hadron structures, 
such as an $s\bar{s}$-quark content of nucleons \cite{Titov97} and reaction mechanisms 
\cite{Mibe05,Kohri10,Sandorfi10}. In order to measure such "complete polarization observables", an essential key technology is to prepare a polarized target with a long relaxation time. We have been developing a frozen-spin polarized Hydrogen-Deuterium (HD) target on the basis of a new technologies in cryogenics, 
superconducting magnet, and HD gas purification. These developments would lead us to the next upgrades of the LEPS experiments \cite{Fujiwara03}. 
An idea for making the frozen-spin polarized HD target came from the work in Refs. \cite{Bloom57,Hardy66}.  
It is suggested that the HD molecule can be used as a frozen spin polarized target \cite{Honig67}.
Many efforts have been devoted for a long time to realize the polarized HD target at Syracuse \cite{Honig89,Honig95}, BNL \cite{Wei00,Wei01,Wei04}, 
and ORSAY \cite{Breuer98,Rouille01,Bassan04,Bouchigny05,Bouchigny09}.
Recently, the HD target has been firstly used for the actual experiment at LEGS \cite{Holbit09},
and will be used at JLab \cite{Sandorfi} and at SPring-8 \cite{Kohri10-1} in near future.

In order to prepare the frozen-spin polarized HD target,
Honig \cite{Honig67} has applied an innovative idea that the HD polarization gradually 
grows up by the spin-flip process between HD molecules and a
small amount of ortho-H$_{2}$ (o-H$_{2}$) with spin J=1, whose aspect was many year ago suggested by Motizuki {\it et al.} \cite{T. Moriya,K. Motizuki}.
Meanwhile after a long period of cooling, o-H$_{2}$ molecules are converted to
para-H$_{2}$ (p-H$_{2}$) with spin J=0. 
If all the o-H$_{2}$ molecules are converted to the p-H$_{2}$ in the HD target, 
the relaxation time of hydrogen polarization becomes very long even at a temperature higher than 4 K.
After an aging process for 2-3 months, the long relaxation time is expected with a hydrogen polarization higher than 84\%. 
The polarization degree is measured by using the NMR method~\cite{PXI-NMR}.
In the conversion process from o-H$_{2}$  to p-H$_{2}$, an important parameter is 
a quantity of remaining o-H$_{2}$ with J=1, which acts as a mediator for spin depolarization when the temperature of the solid HD target rises up to more than 4 K: the lower the impurity of o-H$_{2}$, the longer the relaxation time of hydrogen.
Thus, it is essential to control the admixture of the o-H$_{2}$ in the highly purified HD gas.

When the HD gas is kept in the gas container for a long time, the quality of the HD gas  is deteriorated due to the dissociation process, 2HD $\longleftrightarrow$ H$_{2}$ + D$_{2}$ and the H$_{2}$ and D$_{2}$ gasses are naturally yielded. 
For the HD purification, we installed a distillation still to make the pure HD gas with a purification level of 99.99\%. The next indispensable problem to be solved is how to measure the small admixtures of o-H$_{2}$, p-H$_{2}$, and  D$_{2}$ in the HD gas. The conventional method is to use the quadrupole mass spectrometer (QMS) for gas analysis. 
This instrument makes use of a principle that the analyzed gas is ionized at an ion source by electron bombardment and ions are mass-separated according to the mass/charge  ratio (u/e). 
When the huge HD gas is entered in the ion source of the QMS, small amounts of H$^{+}$ and  D$^{+}$ are produced from HD molecules.
Other ions of H$_{3}^{+}$, [H$_{2}$D]$^{+}$, [HD$_{2}$]$^{+}$, and D$_{3}^{+}$ are also produced by recombination.
This means that D$^{+}$ and [H$_{2}$D]$^{+}$ are misidentified as H$_{2}^{+}$ and D$_{2}^{+}$, respectively.
The QMS cannot distinguish the molecules and fragments with a same mass/charge ratio.  
We call this problem as "fragmentation problem". The fragments produced by ionization are listed in 
Table~\ref{table:fragmentations.table}.

\begin{table}[h!]
  \begin{center}
  \caption{The isomers in the HD gas and expected fragments.}
  \label{table:fragmentations.table}
  \begin{tabular}{ccccccc}
    \hline
     u/e  &  1 &2 &3 &4 &5 &6 \\
    \hline
    \hline
   Molecules &  &H$_{2}^{+}$ &[HD]$^{+}$ &D$_{2}^{+}$ & &   \\
   Fragments   & H$^{+}$&D$^{+}$& H$_{3}^{+}$&[H$_{2}$D]$^{+}$&[HD$_{2}$]$^{+}$&D$_{3}^{+}$   \\
    \hline
 \end{tabular}
 \end{center}
 \end{table}

To overcome the fragmentation problem, we designed a system combined with the gas chromatography and the QMS 
for the HD gas analysis with a high dynamic range of more than 10$^{4}$.
The gas chromatography separates the isomers in the HD gas in terms of time. 
The QMS separates the isomers in terms of mass/change.
By combining two different analysis systems, it is possible for us to analyze 
the H$_2$ and D$_2$ concentrations in the HD gas, precisely.
The James Madison University group has successfully employed a gas chromatography for distinguishing  o-H$_{2}$ and p-H$_{2}$ \cite{JMU}. 
However, a small D$_{2}$ admixture with a level of 0.01\% cannot be clearly observed because an HD long tail conceals a tiny D$_{2}$ peak in the gas chromatography spectrum. 
 
We have developed a new gas analyzer system by combining the gas chromatography 
and the quadrupole mass spectrometer for removing these disadvantages. 
Precise measurements of the o-H$_{2}$ and D$_{2}$ concentrations in the HD gas will be used to effectively polarize the HD target and to obtain the long relaxation time.

\section{System overview}
\begin{figure}[htbp]
  \begin{center}
    \includegraphics[width=80mm]{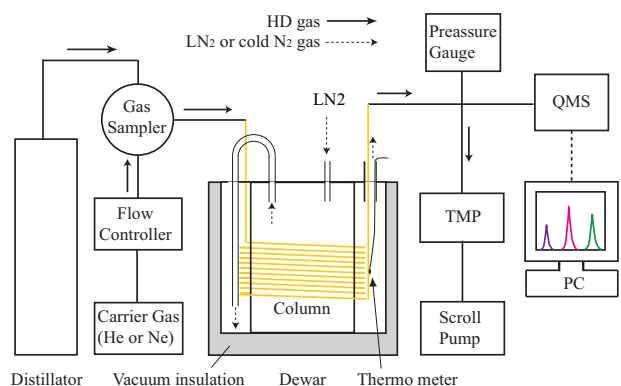}
  \end{center}
  \caption{Schematic diagram of the gas chromatograph and the quadrupole mass spectrometer for the HD gas analysis.  The HD gas is injected with a carrier gas from the entrance of the column via gas sampler. QMS: quadrupole mass spectrometer. TMP: Turbo molecular pump.}  
\label{fig:GC_system.eps}
\end{figure}

Fig. \ref{fig:GC_system.eps} shows a schematic diagram of a new gas analyzer system.
The HD gas with a volume of 500 $\mu$l is infused by a gas sampler into the column. 
The sample gases are pushed in the column together with a carrier gas. 
The isomers of o-H$_{2}$, p-H$_{2}$, HD, and D$_{2}$ in the HD gas are separated by using the difference of adsorbent action to zeolites which are affixed on the inner wall surface of the column.
The separated isomers of the HD gas are analyzed by the quadrupole mass spectrometer as a function of the mass/charge ratio. 
In the usual method of a gas chromatography, the temperature of a thin zeolite column is increased during the analysis because the retention time of the gas is very long. On the other hand, since the hydrogen and deuterium are light and small particles, the retention times of o-H$_{2}$, p-H$_{2}$, HD and D$_{2}$ are as short as a few second at the room temperature.
Because of this, the zeolite column is cooled down to about 110 K to attain a reasonably long retention time for the hydrogen and deuterium gases. Actually, the column is cooled down in a constant temperature dewar by controlling the vapor flow from liquid N$_{2}$ (LN$_{2}$).
The capillary column is installed in the dewar with the vacuum insulation layer. LN$_{2}$ is stored at the center layer, and the cold vapor gas from LN$_{2}$ flows from the inner layer to the outer layer. 
We use the He or Ne gas as a carrier gas. The flow rate of the gas is controlled by using a flow controller. 
The sample which is out of the column is detected by the QMS. This QMS is a product of MKS Instruments ~\cite{MKS} and is specially tuned-up to increase the sensitivity for detecting light molecules such as H$_{2}$, HD, D$_{2}$, and T$_{2}$.

\subsection{Distillator}
To make a polarized HD target with a long relaxation time of more than one month, we need to prepare a highly purified HD gas.
However, if we use highly purified HD gas, the aging time for polarizing the HD becomes long. 
This is an experimental dilemma.
One of methods to overcome this dilemma is to dope  an appropriate amount of o-H$_{2}$ and p-D$_{2}$ impurities in the purified HD gas. 
The appropriate amount is approximately an order of 0.01\%.
The purity of  the commercially available HD gas is about 96\%.
Contaminations are mostly H$_{2}$ ($\sim2$\%) and D$ _{2}$ ($\sim$2\%).

\begin{figure}[htbp]
  \begin{center}
    \includegraphics[width=80mm]{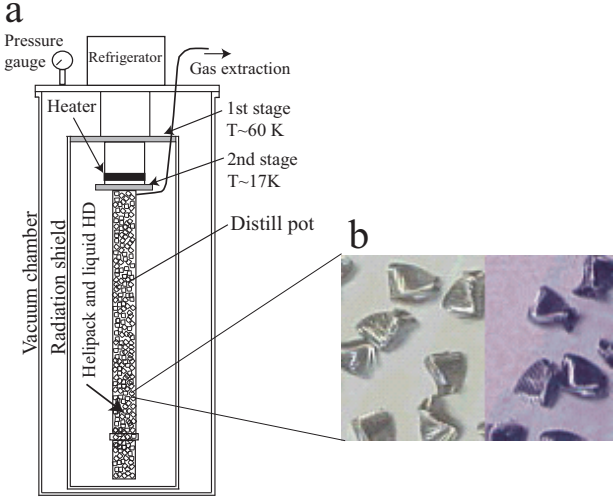}
  \end{center}
  \caption{(a) The design view of the distillator. The H$_{2}$ gas is extracted from the top of the distilllation pot. (b) The photo picture of stainless steel packs which are filled into the pot in the distillator. Each pack has a rolled shape like a coil and twisted for efficient heat conduction between gas and liquid. }  \label{fig:Distill.eps}
\end{figure}

The HD gas is necessary to be purified up to $\ge$ 99.99\% for optimizing the amount of impurities.
The gas distillation system used to purify the HD gas at RCNP is shown in Fig. \ref{fig:Distill.eps}.
The commercial gas is fed to the pot inside the ditillator.
There are stainless cells called "Helipack" inside the pot.
Temperature gradient is realized by cooling the top of the pot and heating the bottom by the thermal radiator. 
Heat exchange between gas and liquid takes place on the cells. 
A gas with a low boiling temperature is extracted from the top of the distillator.
By using the difference of the boiling temperatures of HD (16.6 K), H$_{2}$ (14.0 K) and D$_{2}$ (18.7 K), we separate the HD gas from the others.
In the initial trial, the concentrations of HD, H$_{2}$, and D$_{2}$ were measured with the QMS.
Although the H$_{2}$ and D$_{2}$ concentrations were decreased less than 0.1\%, it was not able to measure the concentrations precisely because of the fragmentation problem.

We operated the distillator for one week to obtain pure HD gas. The commercial HD gas with an amount of 5.3 mol was fed to the pot.
A gas was extracted from the distillator with a flow rate of 2.0 ml/min to the gas storage tanks made of stainless steel.
The H$_{2}$ concentration was very high in the beginning and the HD purity gradually increased in a few days.
After extracting about 1 mol gas, the gas analysis started.

\subsection{Quadrupole mass spectrometer(QMS)}
A quadrupole mass spectrometer (QMS) consists of an ion source, an extraction plate, four cylindrical electrodes, and a Faraday cup. A high mass resolution and compactness are its good features.
The present QMS (made by MKS Instruments, Microvision Plus~\cite{MKS}) is equipped with a multiplier to measure a partial pressure down to $10^{-17}$bar.
The QMS has an ability to well distinguish the masses of He (amu=4.0026) and D$_{2}$ (amu=4.0282), because its performance is customized to measure low mass molecules  such as H$_{2}$, D$_{2}$, and T$_{2}$ in the  mass range of 1$\sim$6~\cite{BouchignyThesis}.

\begin{figure}[htbp]
  \begin{center}
    \includegraphics[width=80mm]{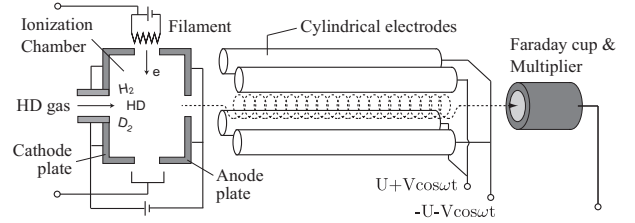}
  \end{center}
  \caption{Structure of the quadrupole mass spectrometer. Constant (U) and alternative (V$\cdot$cos($\omega$t)) voltages are applied to four cylindrical electrodes.
The input HD gas is ionized in the ionization chamber by electron bombardment. Ions are injected along the central symmetry axis of the four cylindrical electrode-rods. An ion satisfying a resonance condition with a frequency $\omega$ makes a helical movement coupled to a displacement movement towards the Faraday cup, and is finally detected with a multiplier.
}
  \label{fig:QMS.eps}
\end{figure}

Constant and alternative voltages are applied to four cylindrical electrodes (quadrupole rods) as shown in Fig.~\ref{fig:QMS.eps}.
The sample gas is ionized in an ionization chamber by an electron impact method and then accelerated by  the extraction plate along to the symmetry axis of the quadrupole rods. Ionized ions move towards the Faraday cup with a spiral motion coupled to a displacement motion.  Ionized ions, which satisfy the resonance condition of the
Mathieu's differential equation, pass the quadrupole rods and reach the Faraday cup. 
The other ions can't pass through the quadrupole rods because of the resonance mismatch. 
The separated ions are detected at the Faraday cup as electric current from a multiplier.
All isomers in the HD gas can be measured by scanning the applied frequency, independently.
We operated the QMS with a pressure of about $10^{-9}$-$10^{-8}$ bar.

\subsection{Gas chromatography(GC)}
We use the Molsieve 5$\r{A}$ PLOT (porous layer open tubular) type column.
The capillary column is made of fused silica and coated by polyimide.
The size of the capillary column is summarized in Table~2.

\begin{figure}[htbp]
  \begin{center}
    \includegraphics[width=80mm]{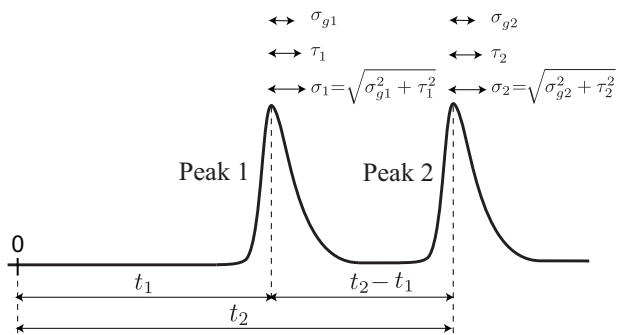}
  \end{center}
  \caption{Expected peaks in the gas chromatography. $t_{1}$ and $t_{2}$ are the retention times. $\tau_{1}$ and $\tau_{2}$ are the time constant of the peak. The separation degree R is defined by Equation \ref{eq:R_peak}. Two peaks are separated completely with R $>$ 1.5.}  \label{fig:peak_example.eps}
\end{figure}

\begin{table}[htbp]
\caption{Specification of the fused silica column. Molsieve 5$\r{A}$ PLOT type column~\cite{Varian} was used. OD and ID stand for the outer and inner diameters of the column.}
\label{fig:column.table}
\begin{center}
\begin{tabular}{cccc}
\hline
OD & ID & Length & Zeolite thickness  \\
  0.70 mm &0.50 mm & 50 m & 0.050 mm \\
\hline
\end{tabular}
\end{center}
\end{table}
Fig.~\ref{fig:peak_example.eps} shows an expected spectrum of gas chromatography as a function of retention time.
The peak in the gas chromatogram data was analyzed by fitting with the
exponential-Gaussian hybrid
function~\cite{Kevin} as follows:
\begin{eqnarray}
f(t)= \left\{\begin{array}{ll}
Hexp \left( \displaystyle \frac{ -(t-t_{R})^{2} }{
2\sigma_{g}+\tau(t-t_{R})} \right), & 2\sigma_{g}+\tau(t-t_{R}) > 0 \\
0,& 2\sigma_{g}+\tau(t-t_{R}) \le 0,\nonumber \\
\label{EGH.eqn}
\end{array}\right.
\\
\end{eqnarray}
where $H$ is the peak height, $\sigma_{g}$ is the standard deviation of
the Gaussian, $\tau$ is the time
constant of the exponential, and $t_{R}$ is the time of the peak maximum.

We define the separation degree R between two peaks as,
\begin{equation}
R = \frac{(t_{2}-t_{1})}{(\sigma_{1}+\sigma_{2})},
\label{eq:R_peak}
\end{equation}
where t$_{1}$ and t$_{2}$ are their retention time, $\sigma_{1}$ and
$\sigma_{2}$ are calculated as,
\begin{equation}
\sigma_{1} = \sqrt{\sigma_{g1}^{2}+\tau_{1}^{2}}, ~\sigma_{2} =
\sqrt{\sigma_{g2}^{2}+\tau_{2}^{2}},
\end{equation}
where $\sigma_{g1}$ and $\sigma_{g2}$ are the standard deviation for the peak 1 and the peak 2, respectively,
and $\tau_{1}$ and $\tau_{2}$ are the time constant for the peak 1 and peak 2, respectively.

Longer column length and longer retention time are considered to give
large separation R.
But very long retention time is not appropriate for the present work.
To measure the HD gas from the distillator efficiently, we have
estimated the best R by changing various parameters such as the temperature of the column and the flow rate of the He carrier gas.

\section{Experimental results and analysis}
\subsection{Effect of temperature and gas flow}
For precise measurement of the concentration of isomers in the HD gas, R is required to be more than 1.5.
To determine the optimum temperature and the flow rate, 
the separation degree R between HD and  D$_{2}$ was measured by changing various parameters of the temperature and the flow rate as shown in Table  \ref{flow_vs_temperature.table}.
Lower temperature  (105K) with a flow rate of 10ml results in the best separation.

\begin{table*}[htbp]
  \begin{center}
  \caption{The separation degree R between HD and D$_{2}$. Errors are the sum of the statistical and systematic errors.} 
  \label{flow_vs_temperature.table}
  \begin{tabular}{ccccccc}
 \hline
 Flow rate   & 1  ml  & 3 ml & 5 ml & 10 ml & 20 ml & 30 ml \\
 \hline
\hline
105 K &  $-$  & $-$  &  5.754$\pm$0.009 &7.988$\pm$0.042 & 4.568$\pm$0.007 & 5.346$\pm$0.008\\
110 K & $-$  & 4.881$\pm$0.005 & 5.346$\pm$0.032 &7.646$\pm$0.017 & 6.107$\pm$0.013 & 6.685$\pm$0.026\\
120 K & 2.610$\pm$0.002 & 4.383$\pm$0.018 & 7.425$\pm$0.054 &7.702$\pm$0.061 & 6.800$\pm$0.043 & 6.862$\pm$0.055\\
130 K &  2.369$\pm$0.002 & 2.416$\pm$0.004 & 3.946$\pm$0.087 &5.253$\pm$0.027 & 3.748$\pm$0.062 & 3.802$\pm$0.016\\
 \hline
  \end{tabular}
 \end{center}
 \end{table*}

\subsection{Measurement of combination with GC and QMS}
\begin{figure}[htbp]
  \begin{center}
    \includegraphics[width=80mm]{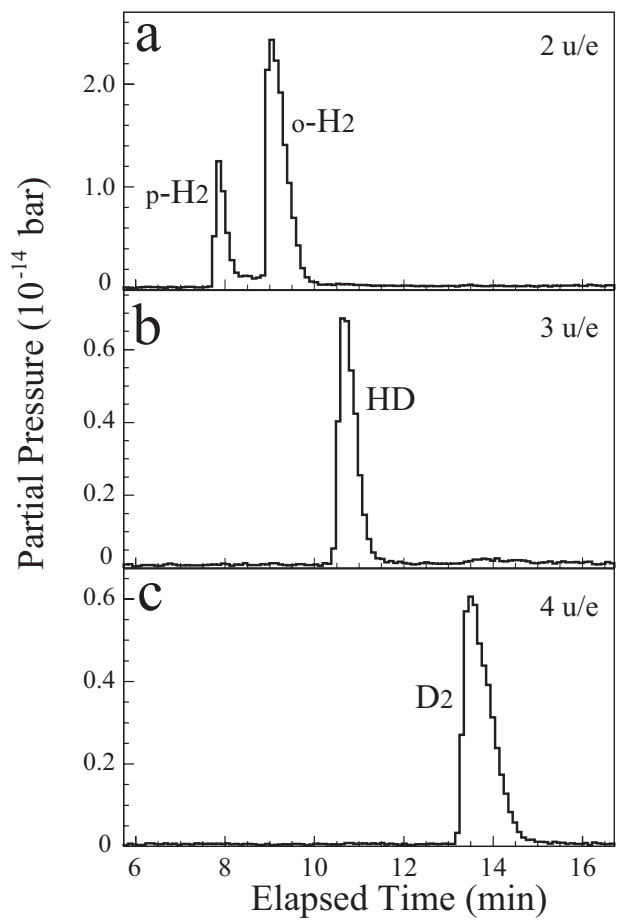}
  \end{center}
  \caption{Gas chromatograms for the gases with the mass/charge ratios of (a) 2  (b) 3 and (c) 4 . 
  The vertical axis is the partial pressure for each gas and the horizontal axis is the elapsed time after the injection of the gas sample.}
  \label{fig:separation.eps}
\end{figure}

\begin{figure}[htbp]
  \begin{center}
    \includegraphics[width=80mm]{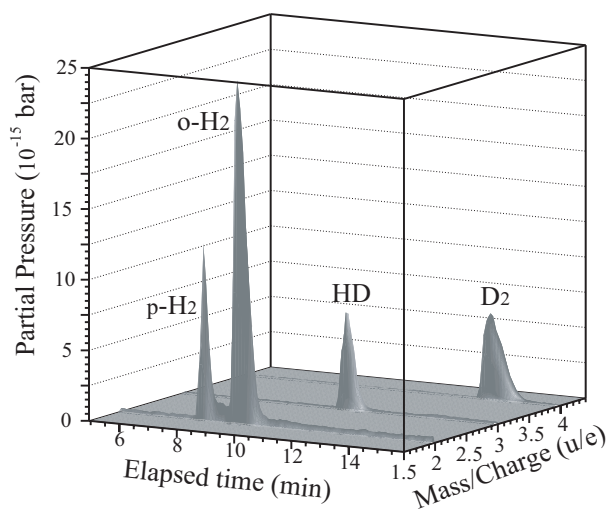}
  \end{center}
  \caption{Three-dimensional plot of the elapsed time of the GC vs the mass/charge ratio measured by the QMS. The z axis is the partial pressure of each gas. }
  \label{fig:gas-rga.eps}
\end{figure}

We prepared a sample gas by mixing appropriate amounts of H$_{2}$, HD, and D$_{2}$ gases.
The sample was analyzed by the gas analyzer system.
The neon gas with a flow rate of 35 ml/min was used to carry the sample
through the column cooled at 105 K.
The p-H$_{2}$ and o-H$_{2}$ were detected at 8 and 9 minutes, respectively,
after the injection of the sample as shown in Fig.~\ref{fig:separation.eps}(a).
The HD and D$_{2}$ gases were detected at 11 and 14 min, respectively,
as shown in Fig.~\ref{fig:separation.eps}(b) and (c).
Using both the GC and the QMS, 
we could observe the p-H$_{2}$, o-H$_{2}$, HD, and D$_{2}$ gases
separately by measuring the retention time in the GC
and by determining the mass/charge ratios in the QMS as shown in Fig.~\ref{fig:gas-rga.eps}.
When the purity of the sample HD gas becomes very high ($\sim$99.9\%), backgrounds disturb clear obsevation of the H$_{2}$ and D$_{2}$ peaks. This problem is discussed in Sec.~\ref{Analysis for pure HD}.

\subsection{Transition between p-H$_{2}$ and o-H$_{2}$}
There are mysterious events between the p-H$_{2}$ and o-H$_{2}$
peaks in the gas chromatograms as shown in Fig.~\ref{fig:separation.eps}(a).
These events are considered to be produced by the transition
from p-H$_{2}$ to o-H$_{2}$ or o-H$_{2}$ to p-H$_{2}$ in
the column at low temperature.
If the p-H$_{2}$ or o-H$_{2}$ molecule interacts with magnetized materials,
the transition can be induced although the materials have not been clearly specified yet.
Such events between the p-H$_{2}$ and o-H$_{2}$ peaks are
observed in another experiment~\cite{JMU}.

\begin{figure}[h!]
  \begin{center}
    \includegraphics[width=70mm]{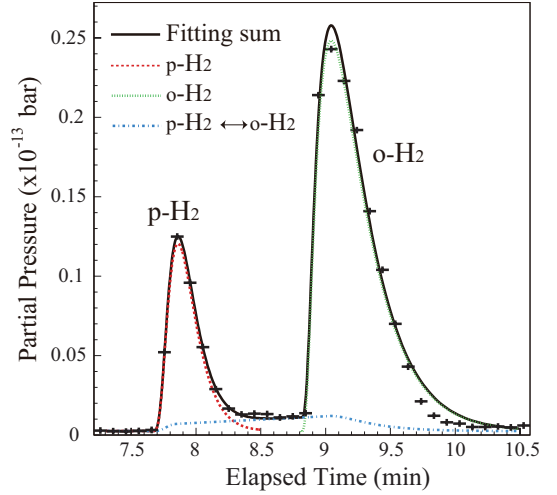}
  \end{center}
  \caption{The fitting to the data with 2 u/e. Peaks of o-H$_{2}$ and p-H$_{2}$ are fit by Eqn.~\ref{EGH.eqn}. Transition region is fit by Eqn.~\ref{EGH_transisiton.eqn}.}  \label{fig:h2fit_kohri.eps}
\end{figure}

We fit the gas chromatogram data as shown in Fig.~\ref{fig:h2fit_kohri.eps} with the function consisting of
two exponential-Gaussian hybrid functions written in Eq. \ref{EGH.eqn},
a function (Eq.~\ref{EGH_transisiton.eqn}) for reproducing the transition events written below,
and a constant for background events.

\begin{eqnarray}
f_{o \Leftrightarrow p}(t)= \nonumber \\
 \left\{ \begin{array}{ll}
CH_{1}exp \left( \displaystyle \frac{ -(t-t_{1})^{2} }{ 2\sigma_{g1}^{2}+\tau_{1}(t-t_{1}) } \right), &\\
 2\sigma_{g1}^{2}+\tau_{1}(t-t_{1})>0& t \le t_{1} \\
\\
C \left(\displaystyle \frac{(H_{1}-H_{2})}{(t_{1}-t_{2})}\times (t-t_{1})+H_{1} \right), &  t_{1} < t < t_{2} \\
\\
CH_{2}exp \left( \displaystyle \frac{ -(t-t_{2})^{2} }{ 2\sigma_{g2}^{2}+\tau_{2}(t-t_{2}) } \right), &\\
 2\sigma_{g2}^{2}+\tau_{2}(t-t_{2})>0& t_{2} \le t ,
\end{array} \right.\nonumber\\
\label{EGH_transisiton.eqn}
\end{eqnarray}  
where $C$ is a normalization factor.
$H_{1}$ and $H_{2}$ are the p-H$_{2}$ and o-H$_{2}$ peak heights,
$\sigma_{g1}$ and $\sigma_{g2}$ are the standard deviation of the Gaussian,
$\tau_{1}$ and $\tau_{2}$ are the time constants, and
$t_{1}$ and $t_{2}$ are the time of the peak maximum, respectively.

The errors for the output of the QMS are estimated from
the measurement fluctuation at a stable pressure.
The reduced $\chi^{2}$ for the fit is 11.3.
The tail of the o-H$_{2}$ peak is not well reproduced, which makes the reduced $\chi^{2}$ large.
The concentrations of p-H$_{2}$, o-H$_{2}$, and transition events
are obtained as shown in Table~\ref{Concentrations of H2}.
The errors of the parameters in the fit and the deviation between the
fit and the data at the o-H$_{2}$ peak tail are considered
as errors of the concentrations.
\begin{table}[h!]
 \caption{Concentrations of p-H$_{2}$, o-H$_{2}$, and transition events obtained by the fit to the data.} \begin{center}
 \label{Concentrations of H2}
  \begin{tabular}{c|ccc}
               & para-H$_{2}$   &  para-H$_{2} \Leftrightarrow$ ortho-H$_{2}$ &  ortho-H$_{2}$  \\
    \hline
  Concentration   &  30$\pm$1\%   & 6$\pm$1\%    & 63$\pm$2\% \\
       \end{tabular}
 \end{center}
\end{table}

\subsection{Analysis for pure HD} 
\label{Analysis for pure HD}
\begin{figure*}[htbp]
\begin{center}
    \includegraphics[width=160mm]{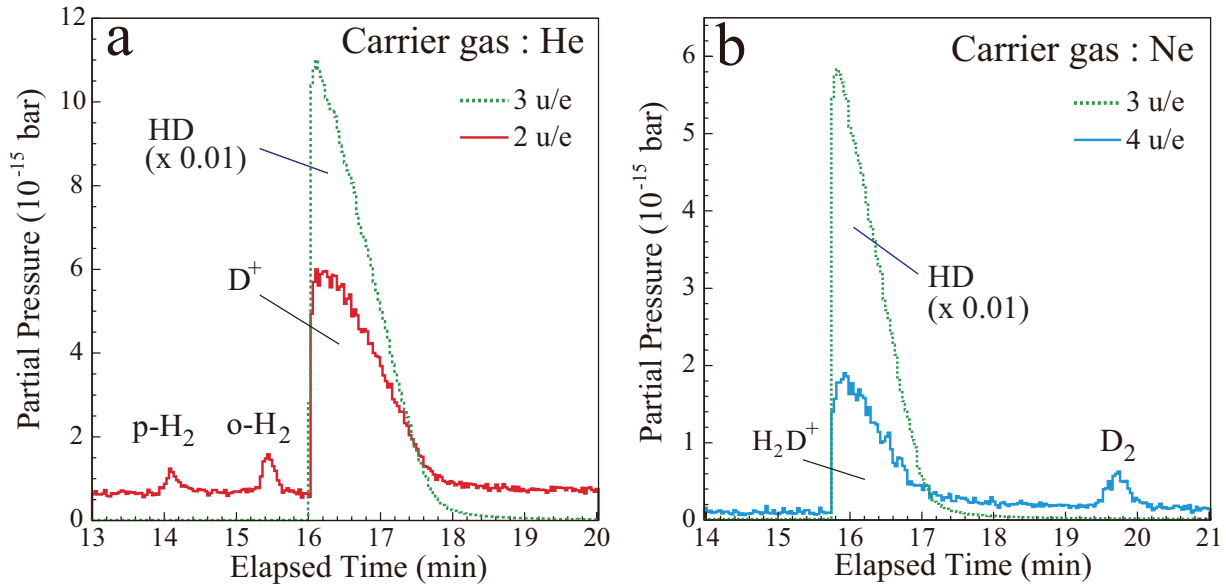}
\end{center}
\caption{(a) Gas chromatograms for a distilled HD gas measured by using the helium carrier gas.
The gas chromatograms with 2 u/e (solid curve) and 3 u/e (dotted curve) are plotted.
(b) Gas chromatogram for the distilled HD gas measured by using the neon carrier gas.
The gas chromatograms with 3 u/e (dotted curve) and 4 u/e (solid curve) are plotted.
The vertical axis is the partial pressure of each gas.
The horizontal axis is the elapsed time. }
\label{fig:H2-HD-D2.eps}
\end{figure*}

The best separation of the peaks was obtained when the helium carrier gas with a flow rate of 10 ml/min was used at 105 K as listed in Table \ref{flow_vs_temperature.table}. 
When the purity of a sample HD gas becomes very high, the experimental condition needs to be optimized in order to reduce backgrounds.
A pure sample HD gas was analyzed by using a helium carrier gas with a flow rate of 1.0 ml/min at 125 K.
As shown in Fig.~\ref{fig:H2-HD-D2.eps}(a), the HD peak is dominantly observed at 16.5 min in 3 u/e.
Another peak is also observed at the same position in 2 u/e.
This peak is due to D$^{+}$ produced from the ionization of HD.
Without the GC, the D$^{+}$ component could not be
easily separated from real H$_{2}$ signals and the measurement
of the H$_{2}$ concentration was uncertain.
The o-H$_{2}$ peak is observed at  15.5 min separately from the
p-H$_{2}$ peak at 14.1 min.
The concentrations of p-H$_{2}$ and o-H$_{2}$ were obtained
as 0.005$\pm$0.001(stat.)$\pm$0.001(syst.)\% and
0.010$\pm$0.001(stat.)$\pm$0.002(syst.)\%, respectively.
The H$_{2}$ events between the p-H$_{2}$ and o-H$_{2}$ peaks
and the tail of the peaks were assumed to be the
p-H$_{2}$ or o-H$_{2}$ with a ratio of 5 (p-H$_{2}$) to 10 (o-H$_{2}$).
Present background level in 2 u/e gas chromatogram enables us to measure the o-H$_{2}$ concentration with a precision better than 0.01\%.

The linearity of the output of the QMS and
the detection efficiency of the GC were taken into account as the systematic errors.
We checked the linearity in the partial pressure region of
10$^{-16}$-10$^{-12}$ bar.
The relation between the amount of sample gas and
the partial pressure measured by the spectrometer were well fitted with a linear function.
The deviation from the linear function was smaller than 12\% of the measured partial pressure.
The detection efficiency was measured by storing the sample gas
and the carrier gas in a tank after passing through the gas chromatography and the QMS.
The uncertainty of the detection efficiency was found to be about 17\%.

The pure sample HD gas was analyzed by using a neon carrier gas
with a flow rate of 1.0 ml/min at 125 K.
As shown in Fig.~\ref{fig:H2-HD-D2.eps}(b), the HD peak is dominantly
observed at 16.0 min in the gas chromatogram with 3 u/e.
At the same location of the HD peak, another peak is
observed in 4 u/e.
This peak is inferred to appear due to the [H$_{2}$D]$^{+}$ molecule produced
by the ionization of HD, which was confirmed by the result
that the [HD$_{2}$]$^{+}$ was also observed at the same location 
in the gas chromatogram with 5 u/e (not shown).
The D$_{2}$ peak is clearly observed at 19.8 min on top of the long tail of
the [H$_{2}$D]$^{+}$ peak.
The concentration of D$_{2}$ in the sample HD gas was
obtained as 0.043$\pm$0.001(stat.)$\pm$0.009(syst.)\%.
Judging from the fluctuation of the background around the D$_2$ peak,  the D$_2$ concentration of 0.01\% can be detected by using the present new gas analyzer system.
\begin{table}[htbp]
\caption{The measured concentrations of p-H$_2$, o-H$_2$, HD, and
D$_2$ in the distilled HD gas}
\label{table:concentration.eps}
\begin{center}
\begin{tabular}{c|c}
Isomers & Concentration \\
\hline
p-H$_{2}$ & 0.005 ~$\pm$0.001(stat.) ~$\pm$0.001(syst.) ~\%\\
o-H$_{2}$ & 0.010 ~$\pm$0.001(stat.) ~$\pm$0.002(syst.) ~\%\\
HD& 99.942~$\pm$0.002(stat.)~$\pm$0.009(syst.)~\%\\
 D$_2$ & 0.043 ~$\pm$0.001(stat.) ~$\pm$0.009(syst.) ~\% \\
\end{tabular}
\end{center}
\end{table}

\begin{figure}[h!]
\begin{center}
\includegraphics[width=80mm]{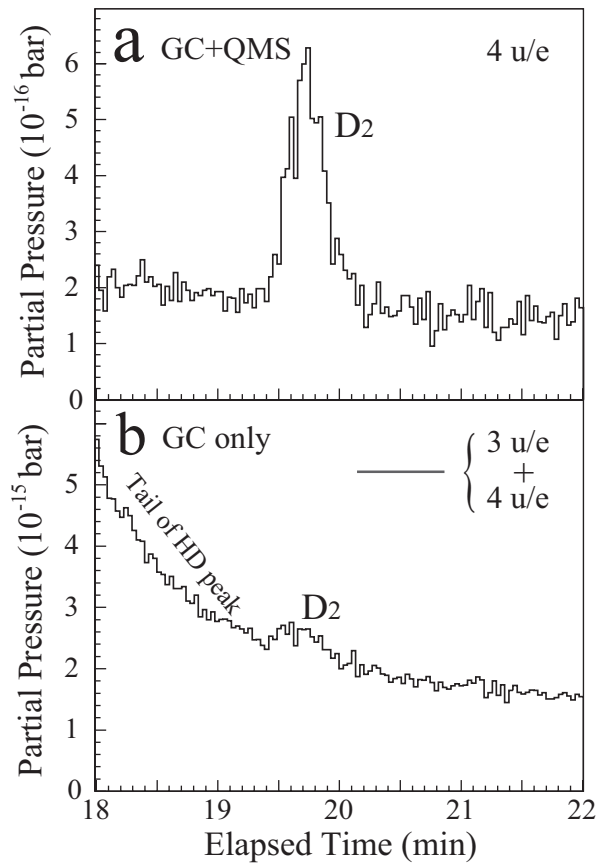}
\end{center}
\caption{(a) The gas chromatogram with 4 u/e for a distilled
HD gas measured by using the neon carrier gas.
(b) The sum of the  gas chromatograms with 3 u/e and 4 u/e for a distilled HD gas measured by using the neon carrier gas.
The vertical axis is the partial pressure of each gas.
The horizontal axis is the elapsed time. }
\label{fig:ontail-v2.eps}
\end{figure}

In Fig.~\ref{fig:ontail-v2.eps}, a D$_{2}$ peak in gas chromatogram with 4 u/e, measured by using the gas chromatography and QMS, is compared with another small D$_{2}$ peak in the gas chromatogram, measured by using the gas chromatography only.
The D$_{2}$ peak is clearly observed with a good S/N ratio in Fig.~\ref{fig:ontail-v2.eps}(a).
However, it is very difficult to find the small D$_{2}$ peak because of
the large background from the tail of the HD peak in Fig. ~\ref{fig:ontail-v2.eps}(b).
The S/N ratio for the D$_{2}$ peak was improved by a factor of 10 in the measurements by using both the GC and the QMS.

\section{Summary}
We have developed a new HD gas analyzer system by combining the gas
chromatography
and the quadrupole mass spectrometer for producing the polarized HD
target to be used for the hadron photoproduction experiments at SPring-8.
The new system enabled us to observe p-H$_{2}$, o-H$_{2}$, HD, and
D$_{2}$ separately.
We succeeded in measuring small concentrations ($\sim$0.01\%) of
p-H$_{2}$, o-H$_{2}$, and D$_{2}$ in the distilled HD gas with good S/N ratios.
Recently, another effort to measure the small concentrations of the
p-H$_{2}$, o-H$_{2}$,
and D$_{2}$ in the HD gas is devoted at JLab~\cite{PrivateCom}.
The JLab group is trying to analyze the HD gas by using the Raman
scattering of laser light.
The accuracy of measuring the concentrations will be improved by
introducing the Raman spectroscopy although the setup is awfully complicate.
It should be noted that the developments of gas analyzing techniques reported in the present work will play an important role in producing the polarized HD target under a well defined HD gas
condition.

\section{Acknowledgments}
The present work is supported in part by the Ministry of
Education, Science, Sports and Culture of Japan and by
the National Science Council of Republic of China (Taiwan).
This work is also supported by Program for Enhancing Systematic
Education in Graduate Schools
in Osaka University.
We thank Dr. T. Kageya for fruitful discussions and Prof. T. Kishimoto
for his encouragement.

\end{document}